\documentclass[11pt]{article}

\usepackage[margin=0.85in]{geometry}
\usepackage{graphicx}
\usepackage{amsmath}
\usepackage{authblk}
\usepackage{lineno}
\usepackage{setspace}
\usepackage{xcolor}
\usepackage[hidelinks]{hyperref}
\usepackage[
    backend=biber,
    style=nature,
    sorting=none, 
    isbn=false, 
    url=false, 
    doi=true
]{biblatex}
\usepackage[most]{tcolorbox}

\newtcolorbox{perspectivebox}[1]{
    colback=blue!5!white,
    colframe=black!80,
    fonttitle=\bfseries\sffamily,
    coltitle=white,
    colbacktitle=black!70,
    title=#1,
    arc=0mm,
    outer arc=0mm,
    left=12pt,
    right=12pt,
    top=12pt,
    bottom=12pt,
    boxrule=0.75pt,
    width=\columnwidth
}

\title{\textbf{Born-Qualified: An Autonomous Framework for Deploying Advanced Energy and Electronic Materials}}

\author[1,2,3*]{Steven R. Spurgeon}
\author[4]{Milad Abolhasani}
\author[1]{Frederick Baddour}
\author[5]{Ryan B. Comes}
\author[6]{Vinayak P. Dravid}
\author[1]{Hilary Egan}
\author[1]{Patrick Emami}
\author[1]{Robert W. Epps}
\author[1]{Davi M. F\'{e}bba}
\author[1]{Renae Gannon}
\author[1]{E. Ashley Gaulding}
\author[7]{Ayana Ghosh}
\author[1,3]{Kenny Gruchalla}
\author[1]{Grace Guinan}
\author[8]{Taro Hitosugi}
\author[9]{Michael Holden}
\author[10]{Sergei V. Kalinin}
\author[11]{Yangang Liang}
\author[1]{John S. Mangum}
\author[11]{Matthew J. Olszta}
\author[12]{Nathaniel H. Park}
\author[1]{Axel Palmstrom}
\author[1]{Michelle A. Smeaton}
\author[1]{Brooks Tellekamp}
\author[1]{Nicholas E. Thornburg}
\author[4]{Raymond R. Unocic}
\author[13]{Daniela Ushizima}
\author[14]{Rama K. Vasudevan}
\author[1]{Robert White}
\author[1]{Andrew Young}
\author[1]{Andriy Zakutayev}
\affil[1]{National Laboratory of the Rockies, Golden, CO, USA}
\affil[2]{Metallurgical and Materials Engineering Department, Colorado School of Mines, Golden, CO, USA}
\affil[3]{University of Colorado--Boulder, Boulder, CO, USA}
\affil[4]{North Carolina State University, Raleigh, NC, USA}
\affil[5]{Department of Materials Science and Engineering, University of Delaware, Newark, DE, USA}
\affil[6]{Northwestern University, Evanston, IL, USA}
\affil[7]{Department of Physics, Indian Institute of Technology Madras, Chennai 600036, India}
\affil[8]{Department of Chemistry, School of Science, The University of Tokyo, Tokyo, Japan}
\affil[9]{Pacific Northwest National Laboratory, Richland, WA, USA}
\affil[10]{University of Tennessee, Knoxville, TN, USA}
\affil[11]{Pacific Northwest National Laboratory, Richland, WA, USA}
\affil[12]{IBM Research-Almaden, San Jose, CA, USA}
\affil[13]{Applied Mathematics and Computational Research Division, Lawrence Berkeley National Laboratory, Berkeley, CA, USA}
\affil[14]{Oak Ridge National Laboratory, Oak Ridge, TN, USA}
\affil[*]{Corresponding author: steven.spurgeon@nlr.gov}

\date{}

\begin{document}

\maketitle

\begin{abstract}

Autonomous science is transforming how we discover materials and chemical systems for advanced energy technologies. However, many initially promising systems never reach deployment. This ``valley of death'' stems from optimization that prioritizes laboratory metrics over industrial viability. We propose a new strategy: ``born-qualified'' autonomous development, which embeds manufacturability, cost, and durability constraints from the outset. This approach is enabled by four pillars, including the development of multi-objective metrics, causal models, a modular infrastructure, and embedding manufacturing in the discovery loop. Realizing this vision will require sustained, community-wide commitment, but the potential return on that investment is commensurate with the scale of the challenge.

\end{abstract}

{\small\textit{Date: \today}}
\clearpage

Modern materials science faces a paradox. While we can now discover materials for advanced energy technologies 10--100 times faster than before, the development time from lab to market has failed to decrease proportionally. Advances in computation now allow us to rapidly screen a vast design space in silico, feeding a growing range of autonomous platforms to synthesize, characterize, and process hundreds to thousands of new compounds.\autocite{Tom.10.1021/acs.chemrev.4c00055} Alongside these computational advances, autonomous materials discovery has accelerated markedly, driven by automated laboratory hardware and Bayesian optimization algorithms that close the loop between prediction and experiment.\autocite{montoyaAutonomousMaterialsResearch2022, chitturiTargetedMaterialsDiscovery2024} However, a widening gap exists between predicting, making, and deploying energy technologies---the ``valley of death.'' This valley emerges not simply because of challenges in cost or durability, which can often be anticipated early in the design cycle, but because of more insidious scale-up and technology-transfer failures: the specialized equipment required to synthesize a promising material at laboratory scale frequently has no industrial analog; synthesis routes that are cost-effective at milligram quantities become economically prohibitive or consume disproportionate fractions of global elemental supply when scaled linearly; and integrating a material into a multilayer or patterned device stack introduces process compatibility constraints that are simply invisible at the single-material discovery stage.\autocite{Kim.10.1038/s41560-025-01720-0} These scale-up constraints are often far more difficult to define and achieve than the material properties that dominate the discovery phase, such as solubility, band gap, or electrical conductivity. Furthermore, the scientific community has long prioritized novelty above other metrics, leading to a proliferation of studies that have failed to translate into actual technologies.\autocite{Trueblood.10.1073/pnas.2401231121} Our ability to solve urgent needs in energy production and resilience depends on addressing this gap. The Autonomous Research for Real-World Science (ARROWS) Workshop,\autocite{nrel_arrows_2025} held in May 2025 in Golden, Colorado, convened leading international experts in autonomous materials science and chemistry to specifically consider these challenges. The central consensus of the workshop was unambiguous: bridging this valley of death is among the most important unsolved problems in autonomous materials science.

Over the past decade, integrated data-driven discovery approaches have proliferated, especially in the field of organic molecular and polymeric material synthesis that lends itself to solution processing methods.\autocite{stachAutonomousExperimentationSystems2021} These advances have coincided with the rise of large language models (LLMs) and high-performance computing, which enable autonomous agents to reason and act in real-time in scientific experiments.\autocite{ramosReviewLargeLanguage2025} Early demonstrations include an organic synthesis robot that can perform chemical reactions and analysis faster than they can be performed manually, leading to discovery of four new reactions.\autocite{grandaControllingOrganicSynthesis2018} Another study showed a 10-fold speedup in the discovery of optoelectronic thin film polymers.\autocite{MacLeod.10.1126/sciadv.aaz8867} Other groups have demonstrated the deployment of mobile robots, compatible with existing laboratory spaces, to conduct molecular photochemical synthesis experiments at scale.\autocite{Dai.10.1038/s41586-024-08173-7} More recent efforts have extended this approach from organic to hybrid and inorganic materials, including nanoparticle, thin film, and polycrystalline powder forms.\autocite{eppsArtificialChemistAutonomous2020, shimizuAutonomousMaterialsSynthesis2020, szymanskiAutonomousLaboratoryAccelerated2023} While powerful, these approaches have largely focused on accelerating the initial discovery phase, while questions of scale-up and manufacturability have largely been deferred to future study.

The result of this narrow focus on discovery is that many candidate materials never deploy. Technical successes accumulate at the bottom of the valley, lacking practical viability. Increasingly, the bottleneck we face is not discovery speed, but rather discovering the right things and developing them fast enough. For energy and electronic technologies, solutions need to scale far more rapidly than the historical pace has allowed. We must optimize autonomous scientific approaches toward deployment from day one.

\section*{The Proxy Problem}

Most autonomous discovery platforms optimize for single key performance indicators (KPIs), such as catalytic activity or ionic conductivity of a material, or sometimes combined figures of merit (FOMs) such as ZT in thermoelectric modules or Baliga\autocite{baligaPowerSemiconductorFOM1989} and Johnson\autocite{johnsonPhysicalLimitationsFrequency1965} FOMs in high-power and radio-frequency electronic devices. While important and easily measured, these KPIs are only proxies for value and not value themselves. These metrics emerge from a research ecosystem where peers, funding agencies, and publishers demand readily comparable benchmarks.\autocite{Trueblood.10.1073/pnas.2401231121} Unfortunately, this ecosystem prioritizes novelty over viability---such as electrocatalysts with record activity under idealized conditions---but can overlook potential catastrophic degradation in real operating environments.\autocite{Kim.10.1038/s41560-025-01720-0} Moreover, even though design based on single KPIs is more computationally straightforward, it is often fundamentally disconnected from deployed value. This problem is exacerbated by discrepancies between experimental and simulated KPIs, as many predictions are based on idealized models (e.g., DFT calculations at 0~K or perfect, defect-free crystal lattices) which fail to capture the effects present in real-world material performance.\autocite{Tom.10.1021/acs.chemrev.4c00055}

Materials and processes are thus optimized for the lab, not the fab. Failure typically occurs late in the development cycle, when manufacturing or cost constraints are finally confronted. Considerable upfront effort then leads to rediscovery of known trade-offs, and promising breakthroughs fail to materialize. A clear example of this is the development of lithium-sulfur batteries in the 2000s.\autocite{Zhu.10.3389/fenrg.2019.00123} Early research into these materials promised a theoretical 1675 mAh g$^{-1}$ specific capacity, 5--10 times higher than lithium-ion batteries, while being abundant and inexpensive. However, these batteries still lack widespread adoption because research has largely focused on specific capacity in coin cells over 50--200 cycles, while overlooking major deployment constraints such as the polysulfide shuttle effect, which causes capacity fade and accelerates beyond 100 cycles. Other considerations, such as integration with the carbon host matrix, binders, and current collectors, lead to a real-world two-fold improvement over lithium-ion batteries and not the promised 10-fold improvement. This challenge is not unique to energy storage; it recurs in photovoltaic technologies (e.g., dye-sensitized solar cells) and in power electronics, where promising wide-bandgap candidates such as diamond have similarly stalled at the laboratory stage.

\begin{figure}[htbp]
\includegraphics[width=\textwidth]{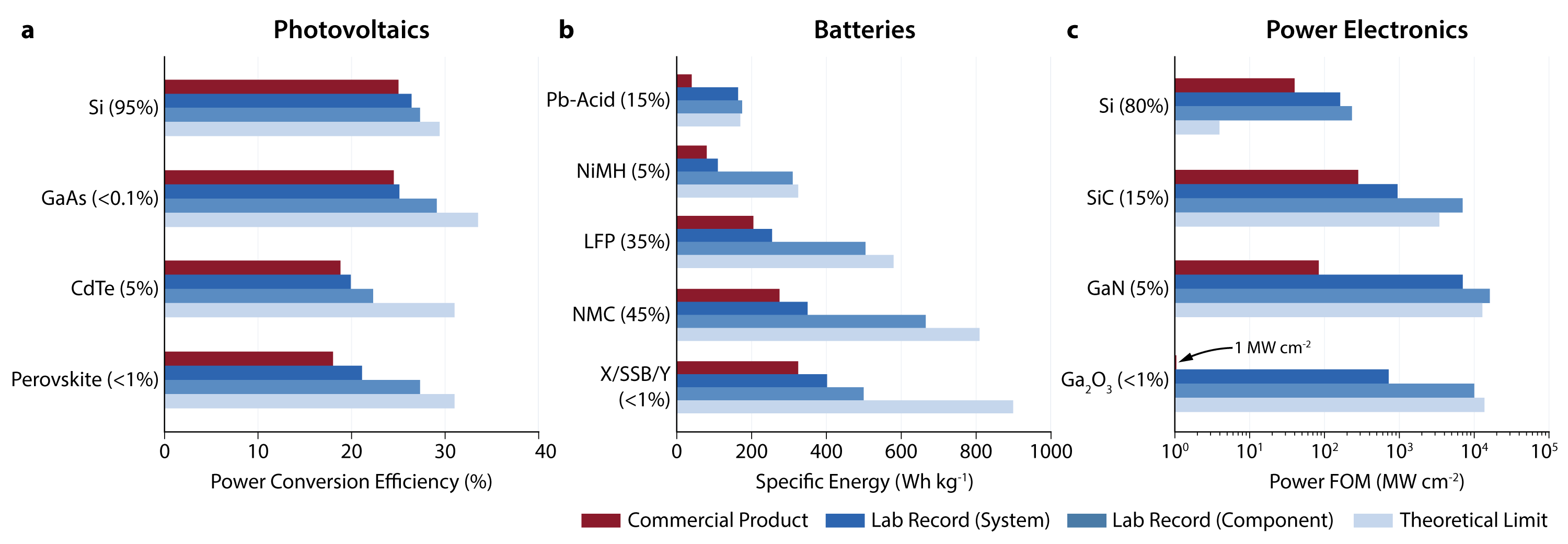}
\caption{\textbf{The Discovery--Deployment Gap in Energy and Electronic Technologies.} 
(\textbf{a}--\textbf{c}) Bars compare commercial product performance, laboratory system and component records, and theoretical limits for key photovoltaic, battery, and power electronic technologies. Percentages indicate estimated market share. X/SSB/Y denotes solid-state battery; electrode chemistry varies by formulation and is detailed in the accompanying data file. NiMH = nickel-metal hydride.}
\label{compare}
\end{figure}

Even if there is no late failure in technology development, the performance gap between laboratory scale records at a sub-component or a component level and commercial products often far exceeds the gap between the theoretical limits and the laboratory record. Fig. \ref{compare} compares the key performance metrics of leading PV, batteries, and power electronic technologies at theoretical, laboratory, and commercial levels, to their corresponding market penetration, organized by the time they have been under development. No matter the application area and how long they have been under development, material technologies with a bigger gap in performance between the laboratory and production tend to occupy a smaller share of the market. This analysis shows how individual material KPIs or even device FOMs may be valuable but still collectively fail to address the complexities of real-world deployment.

For example, while both GaAs and emerging perovskite single-junction PVs show promise in the laboratory, with close to ideal conversion efficiency at the cell and module level, they have not yet been embraced by the market, because of challenges in scale-up reliability and manufacturing cost respectively. On the other hand, the scalable and stable CdTe cell record is not even close to the efficiency limit, yet CdTe modules perform comparably to the cells, resulting in market penetration even under stiff competition from Si. The same holds true for batteries, where lithium iron phosphate (LFP) and lithium nickel manganese cobalt oxide (NMC) technologies and their derivatives together hold 80 percent of the market share, despite not reaching theoretical energy density at the electrode level. For power electronic devices, SiC has higher market penetration than GaN, despite GaN holding higher records than SiC. As shown in Fig. \ref{compare}, the performance of commercial Si devices (80 percent of the market) even exceeds the theoretical limit\autocite{baligaPowerSemiconductorFOM1989} by ``superjunction'' engineering.

\section*{The Born-Qualified Paradigm}

\begin{figure}[htbp]
\includegraphics[width=\textwidth]{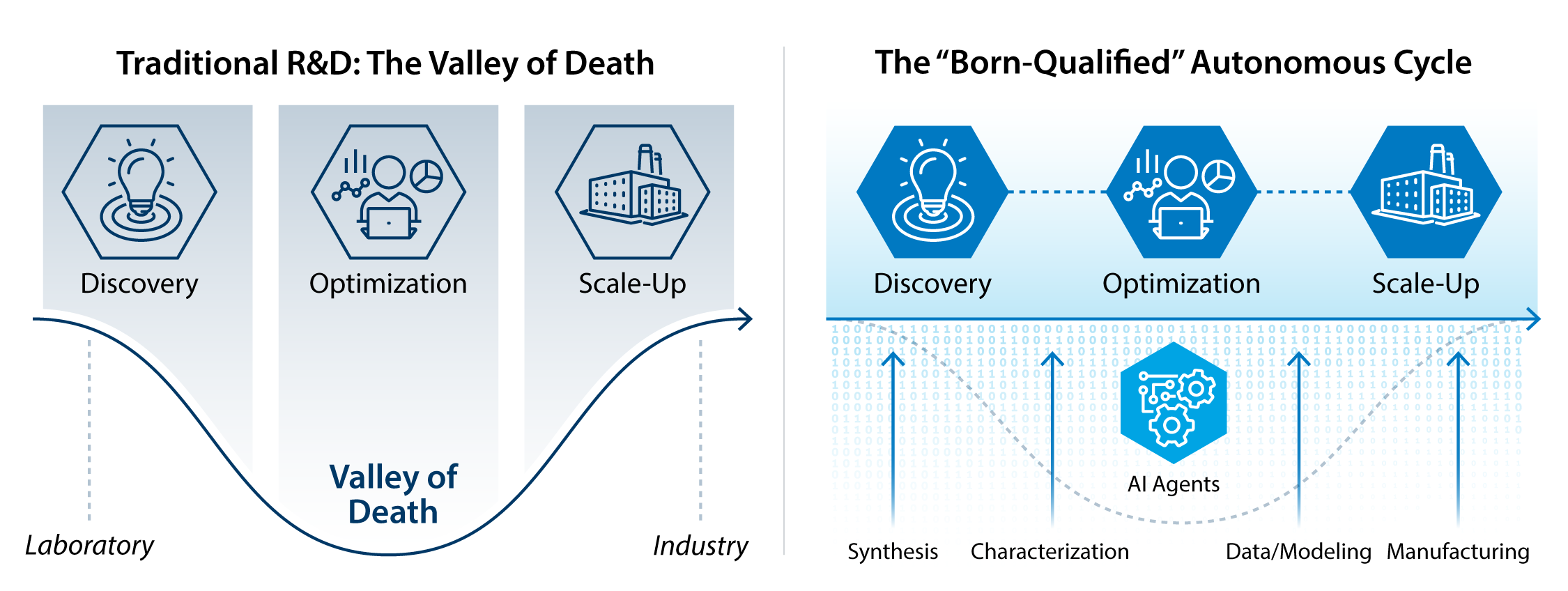}
\caption{\textbf{Traditional Vs. Born-Qualified Discovery.} Comparison of sequential and born-qualified approaches, showing how a linear decision-making process fails to anticipate challenges in scale-up, while a multi-objective approach can more effectively produce feasible designs.}
\label{fig_flow}
\end{figure}

These challenges demand a fundamentally new perspective on autonomous materials discovery. We believe there is an opportunity to anchor discovery in the ``born-qualified'' approach common in additive manufacturing.\autocite{kazmerConcurrentCharacterizationCompressibility2021} This approach refers to materials designed with deployment, manufacturing, and lifetime constraints integrated from inception; it contrasts with earlier efforts, such as the Materials Genome Initiative (MGI), which transformed discovery through a shared infrastructure and high-throughput screening.\autocite{MGI2011} While the MGI dramatically accelerated discovery, manufacturability and deployment were largely treated as downstream issues rather than embedded constraints from the outset. Instead of sequential design---where we first discover, then optimize, then scale a material---we need a more fluid process where all steps inform each other, as shown in Fig. \ref{fig_flow}. Sequential design relies on human-in-the-loop reasoning, forcing binary decisions at each step. While cognitively simple, this approach leads to myopic single-parameter optimization. In contrast, continuous multi-parameter design accounts for all constraints at once and is well suited to autonomous systems that can rapidly evaluate tradeoffs at levels exceeding human cognition. The recently popular co-design approach attempts to address this by jointly optimizing material composition and structure alongside the manufacturing process, tightly coupling computational predictions with experimental validation. However, even this approach is fundamentally limited by siloing of computation-based and experiment-based models, the breadth of experiments (particularly at the pilot scale) typically integrated into the workflows, and a lack of causal understanding of scale-up processes. Here we propose a born-qualified approach that builds on co-design principles to deliver process-informed development of new materials from the outset.

\subsection*{Four Enabling Pillars}

We identify four enabling pillars, summarized in Box 1, that collectively reframe the process of discovery from mere novelty to true deployability: multi-objective metrics, causal models, modular and scalable infrastructure, and manufacturing in the loop.

\begin{perspectivebox}{Box 1 | The Born-Qualified Paradigm}
    \small\sffamily
    \textbf{Born-qualified} refers to materials synthesized with deployment, manufacturing, and lifetime constraints integrated from inception. This approach shifts the focus from laboratory proxies to industrial value through four pillars:
    
    \begin{itemize}
        \item \textbf{Multi-Objective Metrics:} Integrating real-world value functions early in the design cycle.
        \item \textbf{Causal Models:} Using physics-informed, neurosymbolic AI that extrapolates reliably to industrial conditions.
        \item \textbf{Modular, Scalable Infrastructure:} Deploying heterogeneous, reconfigurable hardware architectures managed by AI-driven instrument controllers.
        \item \textbf{Manufacturing in the Loop:} Incorporating pilot-scale processing and realistic manufacturing constraints into the autonomous workflow.
    \end{itemize}
\end{perspectivebox}

\subsection*{1. Multi-Objective Metrics}

Traditional discovery processes are centered on single KPI or FOM optimization. While straightforward and amenable to human cognition, this paradigm poses a serious barrier to deployability. To move beyond this, we must develop and integrate real-world value functions into the design cycle from the outset; these metrics should carry a demonstrable, quantifiable connection to manufacturing outcomes, rather than simply combining multiple laboratory-scale proxies on a Pareto front. This aspiration goes beyond existing multi-objective optimization work, which typically optimizes across several early-stage KPIs without explicitly establishing how those KPIs connect to deployed performance or yield. In contrast to techno-economic analyses that attempt to retrofit early-stage discoveries to complex manufacturing pipelines after the fact, the born-qualified approach embeds those constraints from the start, when they can still influence evolving research trajectories.\autocite{millerRiskMinimizationScaleUp2024} For example, microelectronic design should consider not only switching speed, but also thermal performance, power consumption, and process compatibility with existing fabrication infrastructure. Thermal conductivity and electrical dopability are absolutely critical for power electronic applications, yet neither enters the Baliga nor the Johnson FOM. To meet this need, we must explore the full Pareto front with robust uncertainty quantification, and simultaneously account for uncertainties between experimental measurements and the real-world value functions they are intended to predict. This is particularly important since many deployment-relevant constraints---such as durability under cycling or compatibility with high-throughput epitaxial processes---cannot be measured in situ and must instead be developed through predicted models iteratively validated against scale-up data.

\subsection*{2. Causal Models}

Machine learning has demonstrated remarkable power to reveal correlative trends in data. However, these models struggle to extrapolate to novel, out-of-distribution scenarios. To bridge the gap between discovery and deployment, we require models that embed underlying physical mechanisms to enable extrapolation across conditions, processes, and scales. Combining autonomous experiments that iteratively expand dataset boundaries through active learning with physics-based simulations that ground statistical inference in domain knowledge ensures that uncertainties are informed by both data and physics. Emerging neurosymbolic\autocite{Garcez.10.1007/s10462-023-10448-w} and causal inference\autocite{Ziatdinov.10.1002/adma.202201345} methods can further extend this capability by enabling reasoning beyond initial training data.

Causal reasoning in materials systems remains fundamentally challenging, requiring progression from observational associations to interventional effects to counterfactual inference under partial identifiability constraints. These capabilities are essential for born-qualified autonomous development, where models must predict behavior under deployment-relevant perturbations, such as variations in processing, environment, and scale. Recent advances in causal discovery,\autocite{ghoshCationOrderingDouble2022, kalininAtomicallyResolvedImaging2022, ghoshPhysicsInformedExplainable2024, ghoshMappingCausalPathways2024} intervention-driven learning,\autocite{ghoshCausalReasoningFerroelectrics2026} and causal-relation-informed active learning\autocite{foxActiveCausalLearning2024} establish a foundation for integrating mechanism-aware models into closed-loop workflows, enabling targeted interventions and root-cause analysis. Alongside these data-driven advances, causal discovery frameworks grounded in structural equation modeling and directed acyclic graphs provide complementary tools for extracting process--structure--property relationships from experimental data.\autocite{ziatdinovCausalAnalysisCompeting2020, barakatiCausalDiscoveryLLM2025} Collectively, these approaches allow identification and control of the underlying drivers of performance or failure across scales, facilitating reliable translation from laboratory discovery to deployment.

\subsection*{3. Modular, Scalable Infrastructure}

A key step in implementing real-world autonomous experiments is designing reconfigurable labs for both practical and financial reasons. Academic laboratories and even well-funded companies alike cannot afford to frequently replace specialized research hardware to accommodate new studies. More often than not, there are many existing legacy pieces of equipment that need to be automated. Even for new equipment, the incompatibility between research equipment manufacturers poses a major obstacle to direct integration. Rather than seeking monolithic instrumentation platforms, we must design within heterogeneous and changing laboratory environments. Traditional approaches, where humans manually script low-level interfaces to interconnect such research tools, are highly inefficient. Instead, artificial intelligence (AI), including LLMs that increase the speed and efficiency of coding and act as universal translators between instrument controllers and data formats, should play a key role.\autocite{Fébba.10.1039/d4dd00143e} LLMs can help researchers quickly repurpose instruments and workflows for different experiments, in contrast to the slow, manual processes common today.

\subsection*{4. Manufacturing in the Loop}

Integrating pilot- or prototype-scale processing into autonomous workflows critically impacts deployment. This requires reframing autonomous experiment design from mere discovery of a desired product to discovery of manufacturing routes that actually work.\autocite{Green.10.1063/1.49774879q} This step requires understanding of what processes are industrially relevant, or, better yet, implementing parts of these processes in the research labs: for example, expanding the scope from working solely on electrochemical half-cell anodes/cathodes to integrated full cell batteries, or from stand-alone photovoltaic devices to interconnected solar cell modules. Such a shift demands new levels of partnership between academia and companies, which has been a recommendation of nearly every accelerated materials development workshop of the past two decades. Perhaps the recent emergence of LLMs with deep research modes will make this dream come true, and finally offer academics a window into the industrial manufacturing world, with its practical real-world challenges and pragmatic constraints. Co-developing process models, such as digital twins, that validate and benchmark viability in silico before physical synthesis can dramatically reduce time, cost, and experimental effort, while embedding realistic manufacturing constraints from the outset.

\section*{The Pathway to Implementation}

With these four pillars in mind, we outline a realistic pathway to implementation. Basic research at universities and laboratories has often been siloed from industry, assuming that promising materials would transition well to industrial scale-up. As already argued, this model is incompatible with the pace and return on investment needed for today's advanced energy technologies.

\medskip
\noindent To speed implementation, several key developments are needed:

\medskip
\noindent \textit{Synthesis}. First, we need validated proxy metrics deployable during synthesis: cost-effective, high-throughput diagnostics that reliably predict downstream performance.\autocite{Foadian.10.1021/acsmaterialsau.4c00096} These could be an in-line optical probe to monitor the change in photoluminescence from a semiconductor with the variation in the growth conditions, a non-destructive acoustic measurement of electrode nanoparticle size and agglomeration in a flow reactor, or even a time-series readout from a mass flow controller, power supply, or another component already installed on the instrument. Such proxy measurements are especially helpful when synthesis sequences can be reversed (e.g., in situ etching) or dynamically altered by algorithms on the fly, and the results of such syntheses should be validated with deep, ground truth characterization.

\medskip
\noindent \textit{Characterization}. Second, measurements need to be on timescales and length scales relevant to real-world applications. Autonomous microscopy, for example, can enable rapid, multimodal characterization spanning high-throughput screening at micron-scale and larger---mapping grain boundary networks across device-relevant areas, or quantifying defect distributions across entire wafer regions---down to sub-nanometer structural insight that determines precise atomic positions and bonding environments.\autocite{guinanMindGapBridging2025, guinan3DMXenes2026, Spurgeon.10.1038/s41563-020-00833-z} This capability, demonstrated concretely in recent AI-guided workflows that reconstruct 3D defect topologies across hundreds of thousands of atomic sites,\autocite{guinan3DMXenes2026} bridges the gap between rapid proxy measurements and fundamental structural understanding. While electron microscopy has led the field in AI/ML integration, diffraction and spectroscopy are equally viable modalities for autonomous closed-loop characterization and represent important frontiers for community development.\autocite{unocicSpectroscopy2024} In situ measurements during growth offer a complementary and particularly powerful route to closing the autonomous loop. Reflection high-energy electron diffraction (RHEED) during molecular beam epitaxy (MBE), for instance, provides real-time feedback on surface reconstruction and growth mode that can be directly integrated into autonomous control loops, enabling process optimization at manufacturing-relevant scales.\autocite{comes2022rheed, palmstromFilmCharacterization2023} Similarly, electrochemical battery measurements need to connect initial slow charge/discharge cycles (e.g., solid electrolyte interface formation) to long-term high-rate cycling that evaluates eventual performance and reliability under deployment conditions.

\medskip
\noindent \textit{Modeling and Data}. Third, data from these synthesis and characterization experiments should be standardized and combined with both modeling results and databases of manufacturing constraints into a common, interchangeable library. Modeling methods vary depending on the application, from atomistic simulations of recombination centers in photovoltaic absorbers to molecular dynamics of ion migration in battery electrolytes to finite element analysis of the field distribution and current flow in power electronic devices. Manufacturing constraints can include elemental choices based on supply chain modeling, process rates and thermal budgets from techno-economic analysis, or component assembly guided by recyclability constraints from life-cycle analysis. Ultimately, these data can inform guiding physical mechanisms linking lab-scale synthesis to manufacturing.

\medskip
The scientific community should also embrace new metrics for evaluating novelty, feasibility, and impact. Programs focused on deploying autonomous science, backed by academic and industry consortia, are essential for driving born-qualified design. Just as in other domains of data science, such as the Modified National Institute of Standards and Technology (MNIST) database for comparing optical character recognition (OCR) algorithms,\autocite{lecunGradientbasedLearningApplied1998} or Massive Multitask Language Understanding (MMLU) for benchmarking LLMs,\autocite{hendrycksMeasuringMassiveMultitask2020} we must develop benchmarks for evaluating deployability in autonomous discovery. Such benchmarks will be complex, since they will have to account for both initial discovery and deployed impact, but they will be essential to meaningfully compare the outputs of increasingly opaque AI agents. This will also aid in designing quantifiable value functions for AI agents, which is a gap in the data science community. Researchers fluent in the languages of the domain and data will be able to populate new, open repositories of scale-up attempts and failure, ultimately guiding more intelligent discovery. To illustrate these principles concretely, consider the process of autonomous discovery for power electronics, as shown in Box 2.

\begin{perspectivebox}{Box 2 | Case Study: Autonomous Development of Power Electronics}
    \small\sffamily
    \setlength{\parindent}{1.5em}

\noindent To understand how this new approach may look in practice, we consider the design of a hypothetical ultra-wide band gap semiconductor material for power electronics. In the traditional approach, we aim for novelty: a high critical breakdown field, a large electron mobility, and a high dielectric constant. After years of painstaking screening of new candidate materials and possible synthesis parameters, we arrive at a novel material with a ten-fold improvement in Baliga FOM, leading to a high-impact research publication despite some difficulties in doping and relatively low thermal conductivity that do not enter into the FOM. It is only at this point that we license the technology to a start up for a scale-up towards wafer-scale manufacturing, and quickly find poor substrate scalability leading to low yield of devices, and epitaxial deposition processes incompatible with high-rate manufacturing. Another five years of work are needed to bring it to market, if at all.

\indent Contrast this with the born-qualified approach, where we first invest considerable effort in defining a multi-objective function: an updated power electronic FOM that also mathematically incorporates thermal conductivity and doping efficiency for candidate materials. Next, we work with an industrial partner to consider this updated FOM on a Pareto front with estimated cost per die, taking into account the achievable diameter and defect density of the substrate, as well as the possible epitaxial growth rate for high-thickness drift layers. We then proceed to optimize the composition of another top candidate by alloying, as well as tune manufacturing process variables to improve yield, utilizing statistical microscopy mapping of defects to constrain the possible growth parameter space and influence substrate orientation selection. This process effectively leads to early device prototyping at a die level, resulting in only a three-fold improvement in performance compared to ten. However, this new compound can be manufactured using existing processes and deployed as initial packaged discrete prototypes within 18 months. By traditional scientific measures this might not be considered ``high-impact,'' but it is ready to reshape power electronics today.

\end{perspectivebox}

\section*{Challenges and Realistic Expectations}

Pursuing the born-qualified approach will require confronting several substantial challenges. First, multi-objective optimization is both computationally and experimentally expensive, requiring sophisticated edge or cloud-based computing. Importantly, such computations must be deployed in line with manufacturing prototypes at a comparable latency, or we will not receive feedback in a timely and actionable manner. However, emerging autonomous systems have already shown that they are far better suited to this new paradigm than manual discovery.\autocite{stachAutonomousExperimentationSystems2021} Their greater reproducibility, precision, and standardization make them more amenable to optimization. Likewise, the collective efforts of the community will reduce barriers to entry and lead to repurposable routines for common manufacturing steps, reducing cost and complexity over time.

Second, we must thoughtfully partner with industry to identify manufacturing bottlenecks, develop business cases, and handle proprietary data or trade secrets. While there are many examples of such partnerships, they require a degree of trust and a shared vision for autonomy. Industry has already begun to recognize that the cost of early stage research, when measured against the standard of deployability, leads to wasted time and money on both sides. Emerging privacy-preserving ML approaches can also play an important role here in obfuscating underlying data while allowing models to be collectively trained and disseminated.\autocite{sahaMultifacetedSurveyPrivacy2024}

Third, we must push for a cultural shift in the scientific community. Researchers currently face strong incentives to prioritize novelty, and the metrics by which journals and funding agencies evaluate impact have historically reinforced this tendency. Emerging industry buy-in will help the community understand that deployment is just as important a criterion. Journals and funding agencies, which have already evolved to reward benchmark data sets, architectures, and greater reproducibility, will increasingly see the value in commercial viability as well as scientific novelty.

While there are many potential benefits to the born-qualified approach, it will not be realized without a sustained, community-wide effort. Pilot demonstrations of specific manufacturing processes can take place over the span of 2--3 years, but widespread acceptance may take more time. Nonetheless, it is clear that the current paradigm is no longer sufficient to keep pace with our world's demand for new energy technologies.

\section*{Conclusion}

Autonomous science has already demonstrated that it can substantially accelerate discovery across a broad range of domains. However, we must ensure that these discoveries translate into real-world impact. The born-qualified approach offers a concrete framework to shift our gaze from dazzling novelty to rugged deployability. This shift is urgently needed in the development of advanced energy and electronic technologies, where our ability to deploy new, effective solutions is increasingly important.

Researchers must consider deployment constraints in autonomous system design from the outset. This requires fundamentally rethinking novelty and integrating techniques and optimization strategies that account for the full design cycle. In addition, research will need to establish closer dialogue with industry to better understand manufacturing. By reframing discovery around deployability from the outset, the materials science community is better positioned to translate its considerable capabilities into technologies that address our most pressing energy challenges.

\clearpage
\printbibliography

\clearpage
\section*{Acknowledgments}
S.R.S. would like to thank Nicole Bryant and Harrison Dreves for their help organizing the ARROWS workshop and review. This work was authored in part by the National Laboratory of the Rockies (NLR) for the U.S. Department of Energy (DOE), operated under Contract No. DE-AC36-08GO28308. The views expressed in this article do not necessarily represent the views of the DOE or the U.S. Government. The U.S. Government retains and the publisher, by accepting the article for publication, acknowledges that the U.S. Government retains a nonexclusive, paid-up, irrevocable, worldwide license to publish or reproduce the published form of this work, or allow others to do so, for U.S. Government purposes.

\section*{Funding Statement}
S.R.S., R.G., M.A.S., and A.Z. were supported as part of APEX (A Center for Power Electronics Materials and Manufacturing Exploration), an Energy Frontier Research Center funded by the U.S. Department of Energy, Office of Science, Basic Energy Sciences under Award \#ERW0345 at NLR. R.C. gratefully acknowledges support from the National Science Foundation under DMR-2527684.

\section*{Author Contributions}
S.R.S., H.E., and A.Z. led the writing and conceptualization. All the other co-authors participated in discussions and contributed to the writing and editing.

\section*{Competing Interests}
The authors declare no competing interests.

\section*{Data Availability Statement}
The data used to generate Fig. \ref{compare} is available for download at: \url{https://doi.org/10.6084/m9.figshare.32136955}.

\end{document}